\documentclass[12pt]{article}
\usepackage{cite,psfig}

\def\be{\begin{equation}}
\def\ee{\end{equation}}
\def\bea{\begin{eqnarray}}
\def\eea{\end{eqnarray}}
\newcommand{\mst}[2]{\mbox{\raisebox{-1mm}{$\,\stackrel{#1}{\scriptstyle
#2}\,$}}}
\renewcommand{\phi}{\varphi}

\begin{document}

\thispagestyle{empty}
\vspace*{.5cm}
\noindent
{\large HD-THEP-00-38\hfill August 2000}\\
\vspace*{2.5cm}

\begin{center}
{\Large\bf Natural Quintessence?}
\\[2.5cm]
{\large A. Hebecker and C. Wetterich}\\[.5cm]
{\it Institut f\"ur Theoretische Physik der Universit\"at Heidelberg\\
Philosophenweg 16, D-69120 Heidelberg, Germany}\\[.5cm]
{\small a.hebecker@thphys.uni-heidelberg.de ,\hspace{0.3cm}
c.wetterich@thphys.uni-heidelberg.de}
\\[2.1cm]

{\bf Abstract}\end{center}
\noindent
We formulate conditions for the naturalness of cosmological quintessence
scenarios. The quintessence lagrangian is taken to be the sum of a simple 
exponential potential and a non-canonical kinetic term. This 
parameterization covers most variants of quintessence and makes the 
naturalness conditions particularly transparent. Several ``natural'' scalar 
models lead, for the present cosmological era, to a large fraction of 
homogeneous dark energy density and an acceleration of the scale factor as 
suggested by observation.
\vspace*{2cm}
\newpage

\section{Introduction}
The phenomenology of the expanding universe appears to be converging on a
set of fundamental parameters that includes a non-zero homogeneous energy
density $\epsilon_{vac}\simeq 2\times 10^{-123}\,M_P^4$ (see~\cite{rev} and
refs. therein). Furthermore, large redshift supernova observations suggest
that the cosmological scale factor is accelerating at present~\cite{perl}.
Given the well-known difficulties in explaining why the cosmological
constant should be zero (see, e.g.,~\cite{coco}), it appears to be even
harder to understand a finite number that represents such a tiny fraction
of the natural scale set by the Planck mass $M_P=1.22\times 10^{19}$ GeV.
The scale discrepancies introduced into physics by this large mass have
triggered many attempts and speculations to interpret the Planck mass as
a dynamical scale~\cite{bd}, to understand a possible time evolution of
the cosmological constant~\cite{tdc}, or to have the dynamics of a scalar
field adjusting the cosmological constant to zero~\cite{dol,ba}.
Quintessence as homogeneous dark energy of an evolving scalar field is 
partially successful in explaining a small present-day value of the 
homogeneous energy density~\cite{wet,rp,clw,fj,cds}. It can lead to a 
cosmology consistent with observation~\cite{swz,as,bm,zs,saw,bcn,ams}. 

In this paper, we discuss the naturalness of different realizations of the
quintessence scenario from the perspective of the scalar field lagrangian
\be
{\cal L}(\phi)=\frac{1}{2}\,(\partial\phi)^2\,k^2(\phi)+\exp[-\phi]\,.
\label{la}
\ee
Here and in what follows all quantities are measured in units of the reduced
Planck mass $\overline{M}_P$, i.e., we set $\overline{M}_P^2\equiv M_P^2/(8
\pi)\equiv(8\pi G_N)^{-1}=1$. The lagrangian of Eq.~(\ref{la}) contains
a simple exponential potential $V=\exp[-\phi]$ and a non-standard kinetic
term with $k(\phi)>0$. If one wishes, the kinetic term can be brought
to the canonical form by a change of variables. Introducing the field
\be
\chi=K(\phi) \qquad\mbox{with}\qquad k(\phi)=\frac{\partial K(\phi)}
{\partial\phi}
\ee
one obtains
\be
{\cal L}(\chi)=\frac{1}{2}\,(\partial\chi)^2+\exp[-K^{-1}(\chi)]\,.
\ee
Nevertheless, the important question whether a given quintessence model can
be considered as natural from a field theory perspective can be discussed
particularly simply in terms of the lagrangian of Eq.~(\ref{la}). 

We restrict our discussions to potentials that are monotonic in $\chi$.
(Otherwise, the value of the potential at the minimum must be of the order
of today's cosmological constant, with $V_{min}\approx 10^{-120}$.
Cosmologies of this type are discussed in~\cite{as}.) All monotonic
potentials can be rescaled to the ansatz Eq.~(\ref{la}). An initial value
of $\phi$ in the vicinity of zero corresponds then to an initial scalar
potential energy density of order one. We consider this as a natural
starting point for cosmology in the Planck era. As a condition for
naturalness we postulate that no extremely small parameter should be
present in the Planck era. This means, in particular, that $k(0)$ should be
of order one. Furthermore, this forbids a tuning to many decimal places of 
parameters appearing in $k(\phi)$ or the initial conditions. For natural
quintessence all characteristic mass scales are given by $\overline{M}_P$
in the Planck era. The appearance of small mass scales during later
stages of the cosmological evolution is then a pure consequence of the
age of the universe (and the fact that $V(\phi)$ can be arbitrarily close to
zero). In addition, we find cosmologies where the late time behaviour is
independent of the detailed initial conditions particularly attractive. For
such tracker solutions~\cite{wet,rp,cds} no detailed understanding of the
dynamics in the Planck era is needed. One of our main findings is the
existence of viable cosmological solutions with high present-day
acceleration which are based on functions $k(\phi)$ that always remain
${\cal O}(1)$.

Non-canonical kinetic terms have been considered in cosmology before. For
example, they were used in models of inflation~\cite{sa} and as tool
for the adjustment of the cosmological constant~\cite{ba,rub}, most
recently in the context of quintessence~\cite{hw}. A non-canonical kinetic
term appears in supergravity theories~\cite{bin} and was also used 
in~\cite{ams} to relate the present-day cosmic acceleration to the onset 
of matter domination at $a\simeq 10^{-4}$. In the context of higher 
dimensional unification the identification of $\ln\phi$ with the volume of 
internal space or some appropriate dilaton-type field generically leads to 
a non-canonical kinetic term~\cite{sw}.

It is convenient to analyse the cosmological evolution using the scale
factor $a$ instead of time as the independent variable. In this case, the
evolution of matter and radiation energy density is known explicitly and
one only has to solve the set of the two differential equations for the
homogeneous dark energy density $\rho_\phi$ and the cosmon field $\phi$
\be
\frac{d \ln \rho_\phi}{d \ln a}= -3(1+w_\phi)\,\,\,,\qquad
\frac{d \phi}{d \ln a}=\sqrt{6 \Omega_T/k^2(\phi)}\,\,,
\ee
with $\Omega_T=T/(3H^2)$ the fraction of kinetic field energy and $w_\phi=
p_\phi/\rho_\phi$. Here the cosmon kinetic energy is denoted by $T=
\dot{\phi}^2k^2(\phi)/2$ whereas $p_\phi=T-V$ and $\rho_\phi=T+V$ specify
the equation-of-state of quintessence. Thus, more explicitly, the cosmology
is governed by four equations for the different components of
the energy density $\rho_m,\rho_r,\rho_\phi$ and $\phi$
\bea
\frac{d\ln\rho_m}{d \ln a}=-3\,(1+w_m)\,\,\,,\hspace*{0.8cm}
&&\frac{d\ln\rho_r}{d \ln a}=-3\,(1+w_r)\,\,,
\nonumber\\ \label{fe}\\
\frac{d \ln \rho_\phi}{d \ln a}= -6\left(1-\frac{V(\phi)}{\rho_\phi}
\right)\,\,\,,&&\frac{d\phi}{d \ln a}=\sqrt{\frac{6\,(\rho_\phi-V(\phi))}
{k^2\,(\phi)(\rho_m+\rho_r+\rho_\phi)}}\,\,,\nonumber
\eea
where $w_m=0$ and $w_r=1/3$ for matter and radiation respectively.

For our exponential potential $V=\exp[-\phi]$, the last equation can be 
rewritten as 
\be
\frac{d\ln V}{d \ln a}=-\sqrt{\frac{6\,(\rho_\phi-V)}
{k^2\,(-\ln V)(\rho_m+\rho_r+\rho_\phi)}}\,.\label{veq}
\ee
We note that today's value of $\rho_\phi$ plays the role of $\epsilon
_{vac}$ and $\Omega_\phi=\rho_\phi/(3H^2)$. For a rough
orientation, today's value of $\phi$ must be $\phi_0\simeq 276$ for
all solutions where the present potential energy is of the
order of $\epsilon_{vac}$.

The simplest case, $k(\phi)=k=$ const., corresponds to the original
quintessence model~\cite{wet} with a potential term $\exp[-\chi/k]$. If
$k^2<1/n_b$ (with $n_b=3(1+w_b)$
and $b=r$ for radiation and $b=m$ for matter), then
the scalar field energy $\rho_\phi$ follows the evolution of the background
component $\rho_b$ in a well-known manner. In this case, one finds a
constant dark energy fraction
\be\label{omf}
\Omega_\phi=n_bk^2\,.\ee
This attractor solution can be easily established from Eqs.~(\ref{fe}) and 
(\ref{veq}) by noting 
the constancy of $\rho_\phi/\rho_b$ and $V/\rho_b$. For $k^2>1/n_b$ the 
cosmological attractor is a scalar dominated universe \cite{wet,clw,pli} 
with $H=2 k^2t^{-1},\ w_\phi=1/(3k^2)-1$. 

If a solution obeying (approximately) Eq. (\ref{omf}) is valid during 
nucleosynthesis, the ``right tuning of the clock'' requires $\Omega_\phi
\mst{<}{\sim}0.2$~\cite{wet,bs}. Another constraint arises from structure 
formation since solutions with large constant $\Omega_\phi$ slow down the 
growth of density fluctuations~\cite{fj}. This is described by the simple 
relation~\cite{fj}
\be
\delta_c\sim a^{1-\epsilon/2}\qquad\mbox{with}\qquad \epsilon=\frac{5}{2}
\left(1-\sqrt{1-\frac{24}{25}\Omega_\phi}\right)\,,
\ee
where $\delta_c$ is the density contrast of cold dark matter. The 
formation of galaxies also requires $\Omega_\phi\mst{<}{\sim}0.2$ for a 
sufficiently long time after the onset of matter domination~\footnote{
Our bound is very conservative. A more realistic limit is probably given 
by $\Omega_\phi\mst{<}{\sim}0.1\dots 0.15$.}. 
For building quintessence models,
this constraint is the most stringent one because it requires a recent
increase of $\Omega_\phi$ that is relatively rapid on a cosmological scale. 

It has been emphasized early~\cite{wet} that there is actually no reason
why $k(\phi)$ should be exactly constant and that interesting cosmologies
may arise from variable $k(\phi)$. In particular, one may imagine an
effective transition from small $k$ (small $\Omega_\phi$) in the early
universe (nucleosynthesis etc.) to large $k$ ($\Omega_\phi\simeq 1$)
today~\cite{wet,swz,bcn,mp}.

\section{Leaping kinetic term}\label{leaps}
\begin{figure}[ht]
\begin{center}
\vspace*{.2cm}
\parbox[b]{15.7cm}{\psfig{width=15.7cm,file=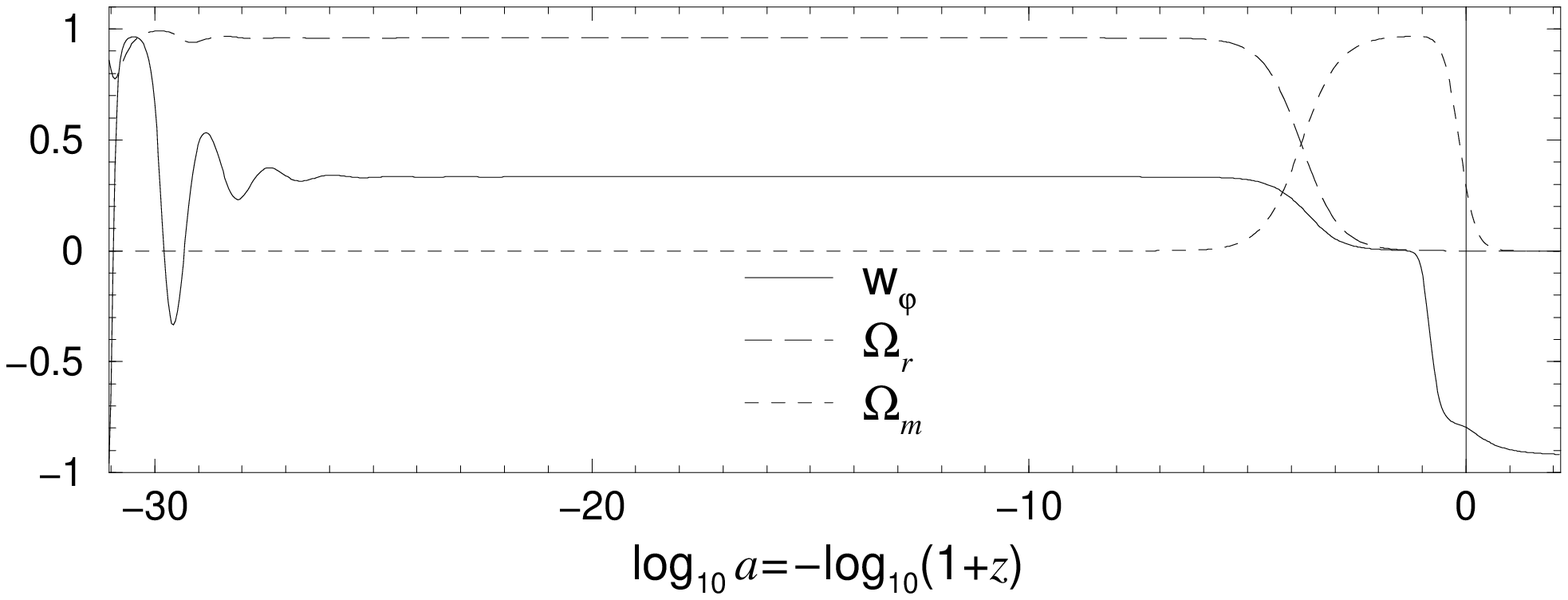}}\\
\end{center}
\refstepcounter{figure}\label{jump}
{\bf Figure \ref{jump}:} Cosmological evolution with a leaping kinetic term.
We show the fraction of energy in radiation ($\Omega_r$) and matter
($\Omega_m$) with $\Omega_\phi=1-\Omega_r-\Omega_m$. The equation of state
of quintessence is specified by $w_\phi$.
\end{figure}

A particularly simple case of a $\phi$ dependent kinetic coefficient
$k(\phi)$ is obtained if $k$ suddenly changes from a small number $k<0.22$
(consistent with nucleosynthesis and structure formation bounds) to a
number above the critical value $1/\sqrt{n_b}$. Consider, for example, the
function
\be
k(\phi)=k_{min}+\mbox{tanh}(\phi-\phi_1)+1\qquad\qquad(\mbox{with}\quad  
k_{min}=0.1\,,\,\,
\,\phi_1=276.6\,)\,,\label{leap}
\ee
that gives rise to the cosmological evolution of Fig.~\ref{jump}. This
model, which completely avoids the explicit use of very large or very small
parameters, realizes all the desired features of quintessence. The 
homogeneous dark energy density tracks below the background component in 
the early universe
($k=0.1$) and then suddenly comes to dominate the evolution when $k$ rises
to a value $k=2.1$ approximately today. With a tuning on the percent level
(the value of $\phi_1$ has to be appropriately adjusted) realistic
present-day values of $\Omega_\phi$ and $w_\phi$ can be realized. In the
above example, one finds $\Omega_{\phi,0}=0.70$ and $w_{\phi,0}=-0.80$.
Note that, due to the extended tracking period, the late cosmology is
completely insensitive to the initial conditions. In the example of
Fig.~\ref{jump}, the evolution starts at the Planck epoch with a total
energy density $\rho_{tot}=1.0$, $\phi=2.0$ and $\dot{\phi}=0$
(corresponding to $\Omega_\phi=0.14)$. We have checked explicitly other
initial conditions, e.g., with $\Omega_\phi$ near one.

The present day value $w_\phi$ can be forced to be even closer to $-1$ if
the leap of $k(\phi)$ is made sharper or the final value of $k$ is made
higher by a simple generalization of Eq.~(\ref{leap}). Thus, all scenarios
between a smoothly rising quintessence contribution and a suddenly emerging
cosmological constant can be realized.

As a limiting case of the sudden increase of $k(\phi)$, one can consider
models where $k(\phi)$ has a singularity at a certain value of $\phi$.
For example, the function
\be
k(\phi)=k_{min}+(\phi-\phi_1)^{-2}\qquad\qquad(\mbox{with}\quad k_{min}= 
0.1\,,\,\,\,\phi_1=277.5\,)\,,
\ee
leads to a cosmology very similar to the one displayed in Fig.~\ref{jump}.
Note, however, that the potential, when rewritten in terms of $\chi=K(\phi)$,
approaches a constant non-zero value at $\chi\to\infty$. Thus, one could
argue that a cosmological constant has, after all, been introduced in a
hidden way. Nevertheless, the lagrangian with non-canonical kinetic term
may open up new perspectives on the problem of sudden cosmic acceleration.
In particular, it appears possible that the sudden rise of the kinetic
coefficient is the result of some transition in the cosmic evolution which
has a natural reason to occur in the present epoch.

\section{Runaway quintessence}\label{runs}
\begin{figure}[ht]
\begin{center}
\vspace*{.2cm}
\parbox[b]{15.7cm}{\psfig{width=15.7cm,file=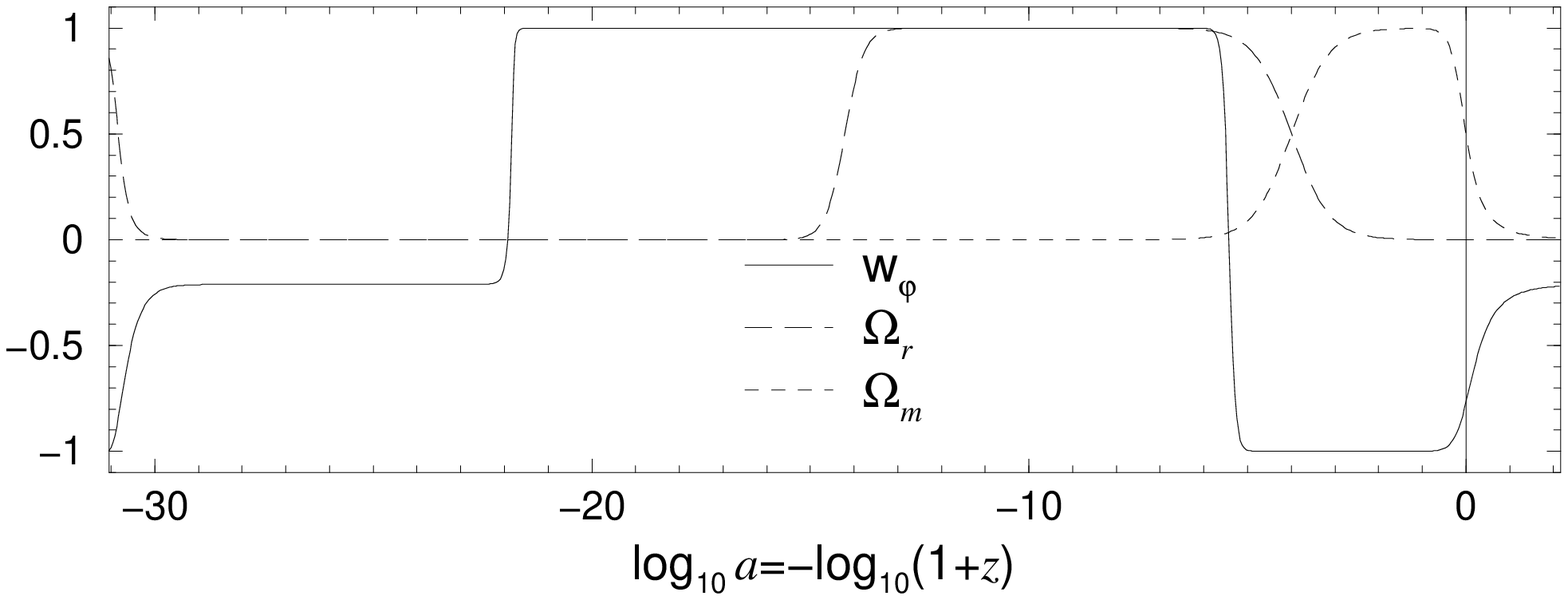}}\\
\end{center}
\refstepcounter{figure}\label{run}
{\bf Figure \ref{run}:} Cosmological evolution with cosmon field running
to very large values and mimicking a small cosmological constant.
\end{figure}

\begin{figure}[ht]
\begin{center}
\vspace*{.2cm}
\parbox[b]{15.7cm}{\psfig{width=15.7cm,file=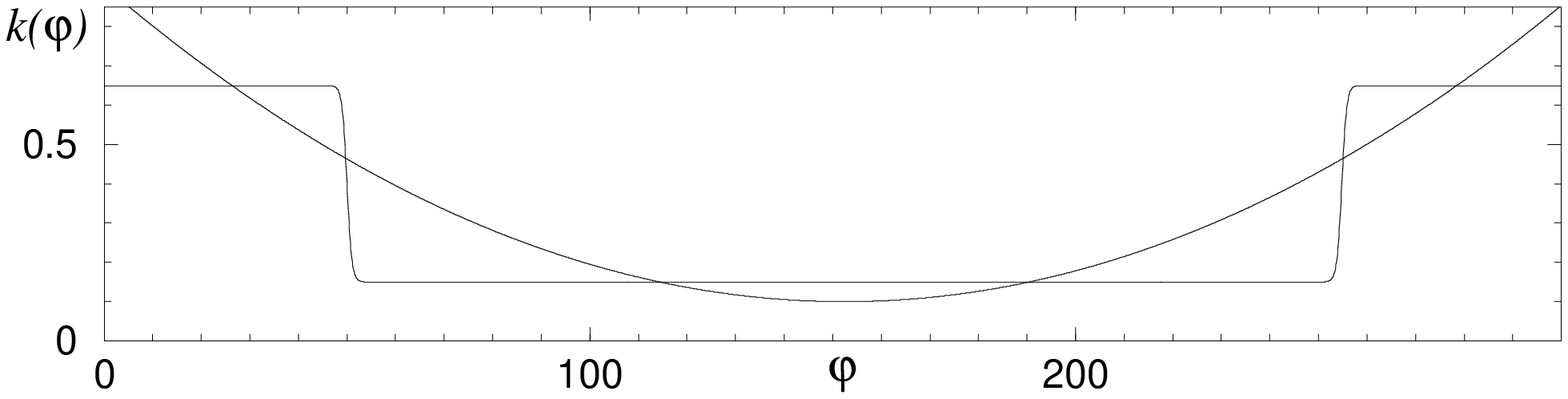}}\\
\end{center}
\refstepcounter{figure}\label{kin}
{\bf Figure \ref{kin}:} Two different kinetic coefficients $k(\phi)$ leading
to the runaway quintessence scenario.
\end{figure}

A somewhat different realization of a cosmology with late-time acceleration
is obtained if the early history of the universe includes a prolonged
period with small $k$. To illustrate this, consider the particularly simple
function
\be
k(\phi)=k_{min}+b\,(\mbox{tanh}(\phi-\phi_1)\,\mbox{tanh}(\phi-\phi_2)+1)
\label{rue}\ee
\[\hspace*{1.5cm}
(\,\mbox{with}\,\, k_{min}=0.15\,,\,\,b=0.25\,,\,\,\phi_1=50.0\,,\,\,
\phi_2=254.8)\,,
\]
which leads to the cosmological evolution of Fig.~\ref{run}. As $\phi$
increases, the coefficient $k(\phi)$ (cf. the almost piecewise constant 
curve in Fig.~\ref{kin}) changes from the large initial value 0.65 to a 
smaller intermediate value 0.15 and back to the large value.

During the initial period with large $k$, the universe inflates and the
background energy density (i.e. radiation) becomes very small. When $k$ 
drops, $\phi$ accelerates and its kinetic energy dominates the universe
($w_\phi=1$). Since
kinetic energy density decays faster than the background component, the
universe becomes radiation dominated after a certain time, from which
point on the value of $\phi$ remains essentially frozen. At this time
$V(\phi)$ has already become very small and the cosmic evolution
(including nucleosynthesis and structure formation) proceeds in the
conventional way. However, at a certain moment (which is chosen to be
approximately now by a moderate tuning of the parameters in 
Eq.~(\ref{rue})) the potential $V(\phi)$ becomes relevant again. Since in 
the meantime $k$ has returned to its large value, a new scalar dominated 
epoch starts. 

In the example of Fig.~\ref{run}, the evolution begins at $\rho_{tot}=
1.0$, $\phi=2.0$ and $\dot{\phi}=0$. These initial conditions lead to
$\Omega_{\phi,0}=0.5$ and $w_{\phi,0}=-0.76$ today. Note that, in this
example, no extended period of tracking exists. Therefore today's
cosmological parameters depend on the initial conditions. However, no
extreme tuning of the lagrangian parameters or the initial conditions is
required. Complete independence of initial conditions could be realized by
introducing a period of tracking before the sequence of events illustrated
in Fig.~\ref{run}.

The same qualitative scenario can also be realized without any abrupt
changes of $k$. For example, the function
\be
k(\phi)=k_{min}+\left(\frac{\phi-\phi_1}{\phi_2}\right)^2\qquad
(\,\mbox{with}\,\, k_{min}=0.1\,,\,\,\phi_1=152.4\,,\,\,\phi_2=170.0)
\ee
(cf. the parabola in Fig.~\ref{kin}) gives rise to a cosmological evolution
that is very similar to the one of Fig.~\ref{run}, if the same initial
conditions at the Planck epoch are used. The essential qualitative feature
of $k(\phi)$ is its small value during an intermediate period, so that
$\phi$ can run away to the large values that correspond to a tiny $V(\phi)$.

\section{Smoothly changing kinetic terms}
In Sect.~\ref{leaps}, late time acceleration was achieved by a relatively
sudden change of the kinetic term. In Sect.~\ref{runs}, it was realized by
an essentially frozen dark energy contribution which suddenly 
becomes the dominant component. Even though the latter does not require
abrupt changes of the kinetic term, the whole cosmic evolution is far from
smooth. In this section, we want to explore whether an interesting
cosmology can be realized with a smooth function $k(\phi)$ and a smooth, 
tracking evolution of $\phi$.

An obvious problem arises from the necessity to produce enough structure in 
our universe. Assuming the approximate validity of the relation 
Eq.~(\ref{omf}) during structure formation, one needs a value $k<0.26$ to 
fulfil the 
condition $\Omega_\phi<0.2$. By contrast, we need $k>0.41$ today to have, 
say, $\Omega_\phi>0.5$. Thus, while $\log_{10}a$ grows by about 3 units or 
less (which is a small fraction of its whole evolution from $\log_{10}a
\simeq-30$ to $\log_{10}a=0)$, a significant change of $k$ has to occur. An 
even stronger rise of $k$ is necessary to account for an appreciable
acceleration of the expansion today. 

Within the approximate validity of Eq.~(\ref{omf}) (with $k\to k(\phi)$) 
one has
\be
\frac{d\,k(\phi)}{d\phi}=\frac{1}{3\sqrt3}\frac{d\sqrt{\Omega_\phi}}
{d\ \ln\ a}\stackrel {>}{\scriptstyle\sim}0.007\,.\label{x}
\ee
Here the bound relates to the time of structure formation and
corresponds to the change of $\Omega_\phi$ from 0.2 to 0.5 mentioned 
before. Together with the value $\phi\simeq 250$ and $k=0.26$ (at the 
borderline allowed for structure formation) this makes it obvious that this 
increase cannot be achieved by a linear rise of $k$ with $\phi$. Cosmologies 
saturating the bound Eq.~(\ref{x}) will not lead to an accelerating universe 
today. 

Let us next consider an exponential form of $k(\phi)$, 
\be
k(\phi)=\exp\left[\frac{(\phi-\phi_1)}{\alpha}\right]\,,\label{ke}
\ee
which allows for a strong growth of $k(\phi)$ during the cosmic evolution. The 
phenomenology arising from this functional form is, in fact, well 
known~\cite{rp,swz} because the corresponding canonical lagrangian possesses 
a simple power-law potential:
\be
V(\chi)=A\chi^{-\alpha}\quad,\quad A=\alpha^\alpha\exp[-\phi_1]\,.
\ee
If $\alpha$ is large, $k$ varies smoothly and $\phi$ follows the growth of 
$\ln a$ : $\Delta\phi\,\,\sim\,\,\Delta(\ln a)$. Thus, if $k$ changed by, 
say, a factor of two between $\log_{10}a=-3$ and today, one may roughly 
expect that it has changed by a factor of $2^{10}\simeq 1000$ since the 
Planck epoch. This violates our naturalness assumption. Smaller values of 
$\alpha$ exacerbate this dilemma. Only for $\alpha\mst{<}{\sim}6$ 
acceleration can be realized~\cite{swz}. In this case $A$ is a very small 
parameter when expressed in units of the Planck mass. In our language this 
situation is highly unnatural\footnote{Even in the case of small $\alpha$ 
no extremely small numbers appear directly in Eq.~(\ref{ke}). The smallness 
of $k(\phi)$ at the initial point $\phi\simeq 0$ arises from the exponential 
factor $\exp[-\frac{\phi_1}{\alpha}]$. This  seems, however, to be only an 
optical improvement.} because the initial value of $k(\phi)$ is very 
small, i.e., $k(0)\approx 10^{-20}$ for $\alpha=6$. 

Let us briefly mention a further interesting aspect of the model with
exponential kinetic coefficient (or, equivalently, the power law potential
for $\chi$).
With an initial condition $\phi\ll\phi_1$, the potential energy $V(\phi)$
is far above the tracking value at the beginning of the evolution
(in~\cite{swz} this is justified by an equipartition requirement).
Therefore, initially $\phi$ runs to very large values (and correspondingly
small $V(\phi)$) as in our runaway-scenario of Sect.~\ref{runs}. If $\alpha$
is very small, the age of the universe is insufficient for $\phi$ to return
to tracking. Thus, the effect of the potential energy in late cosmology is
similar to a cosmological constant: $\phi$ is almost constant and $V(\phi)$
very suddenly becomes the dominant component.
This last possibility, which is quite attractive phenomenologically,
suffers, however, from an ``unnatural'' tiny parameter in the
lagrangian  and a tuning of the initial conditions.
In our opinion, the runaway scenario of Sect.~\ref{runs} represents a
viable alternative with similar characteristics for late cosmology.

Let us finally note that replacing the exponential form of Eq.~(\ref{ke}) by 
a different function, e.g., $k\sim\phi^\beta$, does not solve the problem.
Unless the present era is effectively singled out as in Sect.~\ref{leaps}, 
a rapidly growing $k(\phi)$ implies an unnatural situation in the Planck
era while a relatively flat $k(\phi)$ fails to produce acceleration today.

\section{Conclusions}
We have formulated a condition for a natural quintessence scenario:
in the Planck era no extremely small parameter should appear neither
in the effective action nor in the initial conditions. We have
presented examples that realize this scenario and are consistent
with present cosmological observations of a large fraction in homogeneous
energy density and acceleration in the scale factor. Translating
to a standard kinetic term our examples correspond to a relatively
mild modification of exponential potentials. Some other popular
quintessence scenarios, like the ones based on power law potentials
with moderate powers, do not obey our naturalness criterion.

Despite the consistency of these scenarios with present day observations, 
we feel that two issues are not yet understood in a completely satisfactory 
way. The first one concerns the value of the potential for $\phi\to\infty$.
A modification of the potential into $e^{-\phi}+\lambda$ would introduce an
asymptotic cosmological constant. A tiny value of $\lambda$ is consistent
with cosmological observations but incompatible with our naturalness
criterion. We mention two proposals to answer the question why $\lambda$ 
is precisely zero in the quintessence context. One invokes the dilatation 
anomaly~\cite{wet} and the other is based on a dynamical tuning 
mechanism~\cite{hw} (see also~\cite{quad}). 

The second issue concerns the particular role of the present epoch. In all 
realistic scenarios we have found, the present time is characterized by a 
relatively sharp transition to a scalar dominated universe. Our era is 
singled out by this transition. The question 
``why now'' is much less dramatic than for a cosmological constant: within
quintessence the parameters in $k(\phi)$ have to be tuned on
the percent level in contrast to $10^{-120}$ for the cosmological
constant. Nevertheless, a natural explanation of the special role of
``today'' would be very welcome. We can imagine two solutions to this 
puzzle. Either the present phenomenological constraints weaken such that 
smaller values of $\Omega_\phi$ and $|w_\phi|$ are allowed. Or some 
particularities of the present epoch may affect the dynamics. Some 
proposals are based on the change of the effective equation of state of 
the clustering dark matter at the end of radiation domination~\cite{ams}. 
Another possibility is the coupling of the cosmon to clustering dark 
matter~\cite{wet,ame} that would be ineffective during radiation domination. 

In this context it is worthwhile to recall that the cosmon field $\phi$ 
represents the average value of a fluctuating scalar field in a 
nonequilibrium, inhomogeneous universe. Thus, the potential $V(\phi)$ and 
the kinetic coefficient $k(\phi)$ are, in principle, themselves 
time-dependent dynamical quantities. Their time 
evolution is described by the time-dependent effective action of
nonequilibrium field theory~\cite{wet1} that accounts for the
nonequilibrium values of higher correlation functions. The present epoch
is characterized by the onset of strong nonlinearities in the density
fluctuations and therefore large higher correlation functions. Could this 
be the origin of the effective dynamics discussed in this work?

\end{document}